\documentclass[sigconf]{acmart}

\usepackage{xcolor}
\usepackage{algorithm}
\usepackage[noend]{algpseudocode}

\AtBeginDocument{%
  \providecommand\BibTeX{{%
    \normalfont B\kern-0.5em{\scshape i\kern-0.25em b}\kern-0.8em\TeX}}}

\setcopyright{acmcopyright}
\copyrightyear{2018}
\acmYear{2018}
\acmDOI{10.1145/1122445.1122456}

\begin{document}

\author{Vasyl Pihur}
\affiliation{%
  \institution{Snap, Inc}
  \country{USA}
}
\email{vpihur@snapchat.com}

\author{Scott Thompson}
\affiliation{%
  \institution{Snap, Inc}
  \country{USA}
}
\email{scott.thompson@snapchat.com}

\title{Fuzzy Substring Matching: On-device Fuzzy Friend Search at Snapchat}


\begin{abstract}
About 50\% of all queries on Snapchat app are targeted at finding the right friend to interact with. Since everyone has a unique list of friends and that list is not very large (maximum a few thousand), it makes sense to perform this search locally, on users' devices. In addition, the friend list is already available for other purposes, such as showing the chat feed, and the latency savings can be significant by avoiding a server round-trip call. Historically, we resorted to substring matching, ranking prefix matches at the top of the result list. Introducing the ability to perform fuzzy search on a resource-constrained device and in the environment where typo's are prevalent is both prudent and challenging. In this paper, we describe our efficient and accurate two-step approach to fuzzy search, characterized by a skip-bigram retrieval layer and a novel local Levenshtein distance computation used for final ranking.
\end{abstract}




\maketitle

\section{Introduction}
Snapchat is one of the mainstream social applications with hundreds of millions of active daily users that exchange around 5 billion snaps a day (both photos and videos, not including text messages). Finding the right friends to send this astronomical volume of communications to necessitates convenient searching capabilities. Our universal search system allows users to search for existing friends, new users to friend, lenses, stickers, shows, places, games and many other content types. About 50\% of all searches result in a friend selection, making it a critical point of user search experience. 

Friends list is readily available on the local device to power the app's chat feed and is limited in size to a few thousands. It is also unique to each user. All these considerations, together with the desire to avoid the latency due to the server call, make it a perfect candidate for local, on-device search experience. Historically, we checked for a substring match between the query and the friend's name, despite the obvious high prevalence of typo's \cite{typo} and name ambiguities, such as ``Lucas" vs ``Lukas", for example. Therefore, an efficient search that allows for the approximate name matches is clearly desirable. 

At the core of all fuzzy search approaches lies the idea of a fuzzy or approximate string comparison \cite{approx}. How close is the query string to each of the strings in the corpus? It is usually answered in terms of string edits, such as a total sum of all insertions, deletions and substitutions necessary to transform one string into another. The Levenshtein distance \cite{lev} is a string metric for measuring the difference between two strings in those terms and is widely used for fuzzy search implementations. It corresponds to a very popular global alignment Needleman–Wunsch algorithm\cite{need} which describes exactly which insertions, deletions and substitutions are taking place.

However, the Levenshtein distance, as a metric to compute the distance between the search query and a friend name, is not ideal for our on-device friend search use case. There are two main reasons for this:
\begin{enumerate}
\item It is relatively expensive to compute, taking $O(nm)$ time where $n$ is the length of the search query and $m$ is the length of the friend name.
\item It is a global distance that is not suitable for an interactive search experience where users add one character at a time and may be aiming for a partial string match (for example, ``smi" in ``James Smith").
\end{enumerate}

While the first difficulty with using the Levenshtein distance is well known and, in all fairness, the quadratic time complexity for solving such a complex problem as fuzzy string matching is nothing to sneeze at, the second one deserves more discussion.

If the user enters ``mik" as a search query, as one would guess looking for her ``Mike" friends, computing the Levenshtein distance with the friend corpus that has the following friends in the list [``Mike Petterson", ``Jennifer Mikoilan", ``Mark"] produces some undesirable results. In the first case, the query matches the prefix of the first name and the Levenshtein distance is 12, in the second case, it matches the prefix of the last name and has a distance of 15, while in the last case (where the user entered only their first name when registering on the platform), it does not match much at all and it has a distance of only 2!

What is happening here is the issue of global alignment of the two strings versus the local one. Most of the search queries are not complete in a sense that the user attempted to enter the whole friend name when looking for someone. They enter one character at a time and expect to see friends that ``partially" match the query string. For this specific application, a prefix match is clearly the most desirable and should ideally have a distance of 0. One can make an argument that both ``Mike Petterson" and ``Jennifer Mikoilan" should both have a distance of 0 (no string edits are required for a local, substring match), while ``Mark" should have a distance of 2, since it requires a substitution of ``i" for ``a" and the insertion of ``r".

This idea of local alignment led us to explore the ubiquitous Smith-Waterman sequence alignment algorithm\cite{smith}, widely used in genetics for determining the exact place of a short genetics sequence with respect to a reference genome. It is the right approach, but it does not easily produce a metric, given how the algorithm is defined, so some modifications are necessary for our purposes.

In the following sections, we will describe our solutions to the two challenges that prevented us from using the Levenshtein distance out of the box. The first one solves the computational problem and the second one modifies the Levenshtein distance computation to produce a local alignment score, instead of the global one. In essence, the proposed local Levenshtein distance is related to the Smith-Waterman algorithm in the same way as the original Levenshtein distance is related to the Needleman–Wunsch one. 

\begin{table*}[ht]
\begin{center}
\begin{tabular}{ c c c c c c c c c | c c c c c c c c}
 ' a' & 'ab' & 'bc' & 'c ' & ' 1' & '12' & '2 ' & ' a' & 'a1' & ' b' & 'ac' & 'b ' & 'c1' & ' 2' & '1 ' & '2a' & ' 1' \\ 
 \hline
 1 & 1 & 1 & 1 & 1 & 1 & 1 & 1 & 1 & $\frac{1}{2}$ & $\frac{1}{2}$ & $\frac{1}{2}$ & $\frac{1}{2}$ & $\frac{1}{2}$ & $\frac{1}{2}$ & $\frac{1}{2}$ & $\frac{1}{2}$ 
\end{tabular}
\end{center}
\caption{Skip-bigrams for the string ``abc 12" given $\lambda = \frac{1}{2}$ and $k = 1$. Notice the additional bigrams containing the empty character in the front and the initials at the end. Also, $a1$ is the initials bigram which was appended to the end of the string.}
\label{tab:bigrams}
\end{table*}

\section{Two-step fuzzy search}
In all large-scale search applications or, as in our case, the resource-constrained ones, a rule of thumb is to substitute expensive computations with cheaper ones. The whole approach of retrieval versus ranking in search is based on this idea, where retrieval is a set of cheaper operations that can be performed on all or most elements of the corpus, while the final ranking is usually much more expensive and can be performed on the top-$K$ retrieved results for fine-tuning to produce the final ordered set of relevant results.

So we will use a novel (as far as we know) local version of the Levenshtein distance for ranking and a skip-bigrams based string comparison for the lightweight retrieval layer.

\subsection{Skip-Bigrams Retrieval}
Instead of operating in the string character space, as is commonly done for most fuzzy string comparison algorithms, we will instead compute a distance between two strings that is based on skip-bigrams\cite{skip}. When we talk about bigrams here, we are talking about two characters, not two ``words". For example, the string ``abc" has two bigrams, ``ab" and ``bc". These can be thought of as 0-skip bigrams, as no characters between the first and second characters were skipped. This same string also has a 1-skip bigram ``ac" that skips the character ``b". No higher order skip bigrams are present because of the string length. 

There are at least two desirable properties of the skip-bigrams for our purposes. The first one is that the cardinality of all possible skip-bigrams does not depend on the skip order (its size is still the number of all possible characters squared) and is relatively small in practice, since most character combinations are not observed in a limited corpus. Moving, for example, to trigrams would dramatically increase the number of things we need to track.

However, 1-skip bigram is, to a certain degree, equivalent to a trigram, yet has this second desirable property of being ``fuzzy", meaning that its second character can be arbitrary. Higher order skip-bigrams allow for even more fuzziness. 

Not tracking the position of bigrams within the string allows for significant gains in efficiency, while at the same time offers the ability to accommodate character insertions and deletions. The penalty we pay for this flexibility is that we require a second, more expansive, computation to validate the results.

\subsubsection{Skip-bigramgs Generation}
For each string considered (as well as for the query string), we prepand it with an empty `` " character. The reason for doing that is that we want prefix matches to rank higher than simple substring matches from the middle of the string. 

In addition, since a large portion of our friend names have a display name in the form of ``first\_name last\_name", we would like to allow for the ability to search by their initials \footnote{Thanks to Ben Hollis for the suggestion.}. For example, ``Mike Petterson" should show up in search results for ``mp". For this purpose, we append initials at the end of the original string with a space separator.

Thus, before any bigrams are generated, we modify our input strings with these two modifications. Here are a few examples:
\begin{center}
\begin{verbatim}
'abc'               -> ' abc' 
'abc 123'           -> ' abc 123 a1' 
'mike petterson jr' -> ' mike petterson jr mp'
\end{verbatim}
\end{center}

We then generate bigrams up to the order $k$, where the weight of each bigram is given by 
$$
\lambda^k,
$$
where $\lambda$ is the decay parameter and $k$ is the order of the skip-bigram. In a case when the skip-bigram is present in more than one order, the maximum order is used. Practically, these weights are stored in a hashmap where the keys are the skip-bigrams and values are the corresponding weights.

For example, the string ``abc 12" is converted to the bigrams map shown in Table \ref{tab:bigrams} when $\lambda = \frac{1}{2}$ and $k = 1$. It takes $O(m)$ time complexity to construct the skip-bigram representation of a string where $m$ is the length of the string. Of course, one has to do that for each string in the corpus over which the search is made. This initial, one-time cost associated with this retrieval layer needs to be paid only once, with all subsequent computations having a fixed cost of querying the bigram hashmap.

\subsubsection{Bigrams distance function}
We compute the following distance between two strings based on their bigrams representations
$$
BD(q, s) = \sum_{B_q}(q_i - s_i)^2 - I(q_i = s_i)q_i^2,
$$
where $q$ is the query string, $s$ is the target string, $B_q = {b_1, \dots, b_n}$ is the set of skip-bigrams from the query string, $Q = {q_1, \ldots, q_n}$ are the scores for each skip-bigram from the query string representation, $S = {s_1, \ldots, s_n}$ are the scores for each skip-bigrams from the target string representation and $I(a=b)$ is the identity function equal to 1 if true and 0 otherwise. In a case when a bigram is not present in the target string, its score $s_i = 0$.

As one can see, we penalize mismatches in skip-bigram weights and reward equal weight matches (the more negative the score, the closer the match). This results in an efficient and fast (constant time after construction of the hashmap) distance computation across all target strings, producing a ranked list of results where, by design, the prefix matches are always at the top, substring matches follow next and lastly fuzzy results round up the list.

We could have stopped here and presented the ordered list, but the problem is that the results from this step are more fuzzy than desired, even at relatively conservative distance thresholds. This is due to the fact that the typical query length is less than 5 characters and few spurious bigram collisions across the target string result in a small distance value. For example, ``mik" will match ``m123ik" quite well, despite the fact that three insertions would be necessary and it would be hard to expect the user to make so many typo's.

\subsection{Local Levenshtein distance (LLD)}
Given that our queries are shorter than 5 characters in most cases, even two substitutions, insertions or deletions look too fuzzy to our users with negative engagement consequences. After all, two substitutions out of four characters is already only a 50\% match which is pretty fuzzy. Therefore, ideally we should be showing results that are one insertions, deletion or substitution away, at most.

We already discussed the undesired global properties of the Levenshtein distance and the promise of the Smith-Waterman algorithm to provide a more local match. Combining the two approaches, we propose a new, asymmetric and local Levenshtein metric. We will first describe the Local Levenshtein distance computation in Algorithm 1 and then point out the differences with the original Levenshtein distance.

\begin{algorithm}
\caption{Local Levenshtein Distance (inputs: $q$ - a query string, $s$ - a target string).}\label{lld}
\begin{algorithmic}[1]
\Procedure{LLD}{}
\State n := len(q)
\State m := len(s)
\State d[0..m, 0..n] := 0
\State
\For {$i := 1$ to $n$}  
    \State d[i, 0] := i
\EndFor   
\State
\State minDist := Inf
\For {$i := 1$ to $n$}  
    \For {$j := 1$ to $m$} 
        \If {q[i] = s[j]}
            \State cost := 0
        \Else    
            \State cost := 1
        \EndIf
        \State
        \State top := d[i-1, j] + 1
        \State left := d[i, j-1] + 1
        \State diag := d[i-1, j-1] + cost
        \State d[i, j] := min(top, left, diag)
        \State
        \If {i = n} 
            \State minDist = min(minDist, d[i, j])
        \EndIf    
    \EndFor    
\EndFor  
\State
\State return(minDist)
\EndProcedure
\end{algorithmic}
\end{algorithm}

There are two major differences with the global Levenshtein distance:
\begin{enumerate}
\item We do \emph{not} initialize the first row of the matrix $d$ to $0, \ldots, m$. This idea is borrowed from the Smith-Waterman algorithm and is generally applicable across all local alignment schemes. The reason is that we do not want to penalize for starting a first character match anywhere along the target string.
\item Instead of returning $d[n, m]$, the global score of aligning all characters of the query string with all characters with the target string, we return the minimum of the last row. This, in fact, is the smallest distance of matching all characters of the query string to an arbitrary \emph{sub-sequence} of characters in the target string.
\end{enumerate}

These two small modifications make a huge difference and the metric exhibits behavior that we want: prefix and substring matches have a distance value of 0, while a non-zero score captures the number of modifications required to align the query string anywhere along the target string.

To get a better sense on how local vs global algorithms differ, we compute both in Table \ref{tab:local} and Table \ref{tab:global} for the same query ``mike" and the target string ``hi mcke!". For the global comparison, the minimum is 4 and is found in the bottom right corner of the table, while the local one has a minimum of 1 and it is found on the vertical letter ``e", finishing the match for ``mcke". The result from the local Levenshtein distance indicates that there exists an approximate \emph{substring} match with a single character modification and the exact position of that alignment ends on the second to last character.

One obvious extension here is to have different costs for different substitutions which is influenced by the keyboard layout and would likely need to be language specific. Letter ``a" is often substituted for ``s" in English and should have a smaller cost than any other more distant combinations. This is all complicated by the existence of multiple keyboard layouts in different languages and different methods for entering search queries (typing vs swiping) and so forth. For the first iteration, we set the cost to 1 in all cases. It is possible to also tune the cost of deletions and insertions, perhaps based on the observed user tendencies towards them. This again will remain the focus of future work.

\begin{table}[ht]
\centering
\begin{tabular}{c|cccccccccc}
  \hline
 & & h & i & & m & c & k & e & ! \\
  \hline
  & 0 & 0 & 0 & 0 & 0 & 0 & 0 & 0 & 0 \\ 
m & 1 & 1 & 1 & 1 & 0 & 1 & 1 & 1 & 1 \\ 
i & 2 & 2 & 1 & 2 & 1 & 1 & 2 & 2 & 2 \\ 
k & 3 & 3 & 2 & 2 & 2 & 2 & 1 & 2 & 3 \\ 
e & 4 & 4 & 3 & 3 & 3 & 3 & 2 & \boxed{1} & 2 \\ 
\end{tabular}
\caption{Local Levenshtein distance between query ``mike" and target string ``hi mcke!". Notice the first row being all 0's and the minimum distance of 1 occurring at row 5 and column 8, indicating a single substitution of ``i" for ``c".}
\label{tab:local}
\end{table}

\begin{table}[ht]
\centering
\begin{tabular}{c|cccccccccc}
  \hline
 & & h & i & & m & c & k & e & ! \\
  \hline
  &  0&	1&	2&	3&	4&	5&	6&	7&	8\\
m&	1&	2&	3&	4&	3&	4&	5&	6&	7\\
i&	2&	3&	2&	3&	4&	3&	4&	5&	6\\
k&	3&	4&	3&	4&	5&	4&	3&	4&	5\\
e&	4&	5&	4&	5&	6&	5&	4&	3&	\boxed{4} \\
\end{tabular}
\caption{Global Levenshtein distance between query ``mike" and target string ``hi mcke!". This is just to contrast this with the local version. Notice that the distance is equal to 4 (bottom right) and both the first row and the first column are filled with non-zero counts.}
\label{tab:global}
\end{table}

\subsection{Combining retrieval and Local Levenshtein distance}
In the offline phase, we pre-compute all skip-bigrams for all target strings in the corpus and save them in hashmaps, one hashmap for one target string (Algorithm \ref{build}).

\begin{algorithm}
\caption{Offline: Build bigrams hashmaps for the $K$ strings in the corpus ($k$ - a skip-bigram order, $\lambda$ - a decay factor, a set of $K$ documents with strings $s_1, \ldots, s_K$).}\label{build}
\begin{algorithmic}[1]
\Procedure{Build Bigrams Hashmaps}{}
\For {$i := 1$ to $K$} 
    \State $M_i$ := Bigrams($s_i$, k, $\lambda$)
\EndFor    
\State return($M$)
\EndProcedure
\end{algorithmic}
\end{algorithm}

In the online (querying) phase, we then perform Algorithm \ref{fs} where we loop over all target strings in the corpus and output strings with Local Levenshtein distance $t <= 2$.

\begin{algorithm}
\caption{Online: Fuzzy Search (inputs: $q$ - a query string, $k$ - a skip-bigram order, $\lambda$ - a decay factor, a set of $K$ hashmaps built offline by Algorithm \ref{build}.}\label{fs}
\begin{algorithmic}[1]
\Procedure{Fuzzy Search}{}
\State queryBigrams := Bigrams(q, k, $\lambda$)
\State finalResults := []
\For {$i := 1$ to $K$} 
    \State bd := BD(queryBigrams, $M_i$)
    \If {bd <= $t_1$}
        \State lld := LLD(q, $s_i$)
        \If {lld <= $t_2$} 
            \State finalResults := append(finalResults, $s_i$)
        \EndIf    
    \EndIf 
\EndFor  
\State    
\State return(sort(finalResults))
\EndProcedure
\end{algorithmic}
\end{algorithm}

We are currently experimenting with different parameter configurations. However, we have settled on both $t_1$ and $t_2$ to be equal to 1 and the $\lambda$ parameter is preferred to be 1, as well. The argument for the skip order $k$ is still being investigated, but in practice it looks like 1 will also be sufficient (we do not observe a higher degree of friend discoverability with large skip numbers). We also found that enabling fuzzy capabilities for search queries with one or two characters is not necessary, as it introduces too many noisy results even with a Levenshtein distance of 1 (every string matches for a single character in this case and a single character match is sufficient for the two character case which is too fuzzy for practical applications).

\section{Discussion} \label{sec:discussion}
When we embarked on the fuzzy search capabilities, the number one issue we needed to solve was performance. Searching over several thousands of friends on a low-end Android device after each character that the user types is expensive and any proposed solution would have to be reasonably light. We are pleased to report that we have not encountered any performance issues after many months of experimentation. In addition, there are several major optimization that could be done to further increase the efficiency. For example, a C++ implementation would be still much faster, or an incremental local Levenshtein distance computation could be implemented where each new input character just adds a new row to Table \ref{tab:local}, instead of recomputing the whole table after each character. There are, of course, other known optimizations for computing the Levenshtein distance itself \cite{improv}.

We are still experimenting with the final ranking, which is currently based on the local Levenshtein distance first and then the bigrams distance second. With $\lambda = 1$, this puts prefix matches first, substring matches second and then fuzzy matches last. Of course other factors, such as how recently you have interacted with a particular friend, are also important and need to be incorporated into the final search ranking through well-known ranking techniques \cite{ltr}. 

The proposed approach to fuzzy search has been specifically designed with a small corpus in mind (at most several tens of thousands items), but more importantly where ``documents" in this corpus are very short. Here, the name of each friend is likely under two dozen characters and often much shorter. This is obviously key to the efficiency aspect of the bigrams algorithm which loses its discriminating abilities as the number of bigrams in a target string grows large. We specifically avoided any form of tokenization because it was not necessary for our application. For much larger applications and more complex documents, other approaches may prove more practical \cite{set}.


\bibliographystyle{ACM-Reference-Format}
\bibliography{sample-base}

\end{document}